\documentclass[twocolumn]{aastex63}

\usepackage{hyperref}
\usepackage{amsmath}

\shorttitle{GW190521 From the Merger of Ultra-Dwarf Galaxies}
\shortauthors{}

\def\solm{M$_{\odot}\,$}
\def\msol{M$_{\odot}\,$}

\def\kms{km s$^{-1}$}

\def\solm{M$_{\odot}\,$}

\def\kms{km s$^{-1}$}

\def\casgm20{CAS-G-M$_{20}\,$}
\def\m20{M$_{20}\,$}

 \def\kms{\,km\,s$^{-1}$}

\begin{document}

\title{GW190521 from the Merger of Ultra-Dwarf Galaxies}

\author[0000-0002-6011-0530]{Antonella Palmese}
\affiliation{Fermi National Accelerator Laboratory, P. O. Box 500, Batavia, IL 60510, USA}
\affiliation{Kavli Institute for Cosmological Physics, University of Chicago, Chicago, IL 60637, USA}
\correspondingauthor{Antonella Palmese}
\email{palmese@fnal.gov}

\author[0000-0002-0786-7307]{Christopher J. Conselice}
\affiliation{Jodrell Bank Centre for Astrophysics, University of Manchester, Oxford Road, Manchester UK}



\begin{abstract}

We present an alternative formation scenario for the gravitational wave event GW190521, that can be explained as the merger of central black holes from two ultra--dwarf galaxies of stellar mass $\sim 10^5-10^6 ~M_\odot$, which had themselves previously undergone a merger. The GW190521 components' masses of $85^{+21}_{-14}M_\odot$ and $66^{+17}_{-18}M_\odot$ challenge standard stellar evolution models, as they fall in the so--called mass gap. We demonstrate that the merger history of ultra-dwarf galaxies at high redshifts ($1\lesssim z \lesssim 2$) matches well the LIGO/Virgo inferred merger rate for black holes within the mass range of the GW190521 components, resulting in a likely time delay of $\lesssim 4$ Gyr considering the redshift of this event. We further demonstrate that the predicted time-scales are consistent with expectations for central black hole mergers, although with large uncertainties due to the lack of high--resolution simulations in low--mass dwarf galaxies. Our findings show that this black hole production and merging channel is viable and extremely interesting as a new way to explore galaxies' black hole seeds and galaxy formation. We recommend this scenario be investigated in detail with simulations and observations.

\end{abstract}

\keywords{Gravitational Waves; Galaxy Mergers; Dwarf galaxies;  Intermediate-mass black holes; Supermassive black holes; Galaxy evolution}

\reportnum{FERMILAB-PUB-20-482-AE}


\section{Introduction} \label{sec:intro}

The sources and channels that can potentially produce the gravitational wave (GW) events detected to date with LIGO/Virgo pose interesting astrophysical questions. These include understanding the origin of black holes and other compact objects, as well as a wide range of fundamental physics and cosmology problems that can be address by analyzing such events. Since the discovery of gravitational wave events produced by black holes \citep{GW150914}, a major question has been how these massive stellar black holes were formed, and how they could produce binary systems able to merge within a Hubble time. Through analyses of the detected GW signals, LIGO/Virgo are currently able to distinguish between different types of sources. This has resulted in the identification of black hole--black hole (BH) mergers, mergers of BHs and other compact objects such as neutron stars, and binary neutron stars mergers such as GW170817 \citep{ligobns}, the first gravitational wave source with an electromagnetic counterpart \citep{GBM:2017lvd}. The origin of the massive stellar black hole mergers however is still up for debate.  

On May 21st 2019, the LIGO/Virgo Collaboration alerted the astronomical community of a new gravitational wave event from a compact object merger \citep{GCN1,GCN2}. Recently, this event has been confirmed to be the result of the coalescence of two black holes with masses of $85^{+21}_{-14}M_\odot$ and $66^{+17}_{-18}M_\odot$ (90\% Credible Interval, CI) \citep{GW190521,190521_properties}, further challenging stellar evolution theories to explain the origin of these black holes.

GW190521 has a final mass of 142$^{+28}_{-16}$ \msol, which places its remnant at the lower end of the so-called intermediate mass black holes (IMBHs). The existence of these black holes has long been sought, although their origin is still elusive. In the Milky Way, several solar mass black holes have been identified through X-ray binaries, and the existence of central supermassive black holes (SMBH), that can be as massive as several times 10$^{9}$ \solm, in galaxies, has been verified through a variety of methods. The origin of black holes between these extremes is however still unclear, although it could provide interesting ground to also explain the formation of SMBHs. The LIGO/Virgo events from the first two observing runs were more massive than the BHs in X-ray binaries, with components masses of a few tens of solar masses \citep{gwtc}.  While black holes in this mass range had not been observed before, it is possible that these systems could have been formed from massive stars in metal poor, and presumably distant, star formation events (e.g. \citealt{2016Natur.534..512B,limongi}).

However, GW190521 does not easily fit into this picture as the masses of the two merging systems (and more significantly the mass of the primary) fall in the high ``mass gap'', corresponding to the range between $\sim 65$ and $135M_\odot$. The expectation of stellar evolutionary models is that pulsational pair instability (PPI) and pair instability supernova (PISN) prevents the formation of remnant black holes above $\sim 65 M_\odot$ from stars with helium cores of mass $\sim32-64~M_\odot$ and $\sim64-135~M_\odot$, while higher mass stars ($\gtrsim 200~M_\odot$) produced in low-metallicity environments can form BHs with $\gtrsim 135~M_\odot$ through direct collapse (e.g. \citealt{Barkat,Fryer_2001,Farmer_2019}).  Therefore, alternative channels for black hole merger production could provide more plausible scenarios.  It is worth looking at other possibilities and some of these could come from galaxy evolution processes such as galaxy mergers.  

While works have shown that it is possible to form black holes such as those in GW190521 through stellar evolution (e.g. \citealt{farrell2020gw190521,kinugawa2020formation}), alternative theories have proposed that the LIGO/Virgo compact objects could be explained with Primordial Black Holes (PBH, \citealt{Carr:1974nx}; e.g. \citealt{Clesse:2016ajp,tsai2020gw170817}). GW190521 could contain PBHs only if they can significantly accrete mass before reionization (\citealt{pbh_190521}). Another interesting scenario for the formation of compact object binaries is through dynamical interactions in dense stellar environments (e.g. \citealt{Portegies_Zwart_2000,Coleman,palmese17,rodriguez19}), and through assisted inspiral in AGN disks (e.g. \citealt{McKernan12,Bartos_2017}). Given the properties of the binary and the inferred rate of GW190521-like events, \citet{190521_properties} do not find strong evidence for any of these scenarios to be favored, although \citet{romeroshaw2020gw190521,gayathri2020gw190521,fragione2020origin} argue that gravitational capture could explain this event. Beyond the Standard Model physics could also produce GW190521--like stellar black holes \citep{sakstein2020standard}. On the other hand, \citet{fishbach2020dont} show that, through the use of an informative mass prior, the binary could be composed by a BH below the mass gap and a BH with mass above it. 

Another possibility, first proposed in \citet{Conselice_20}, is that the black holes detected by LIGO/Virgo are produced at the centers of ultra-dwarf galaxies.  As explained in \citet{Conselice_20}, ultra-dwarfs are low-mass galaxies which are potential analogs of the faint dwarfs studied in the Local Group and nearby universe.   The argument consists in an extrapolation of the well-known central black hole mass -- galaxy mass relation that has been measured for galaxies down to stellar masses of $M_\star \sim 10^{8}$ \solm.  Whilst we do not have observations of central black holes in lower mass galaxies, there is an increasingly large amount of evidence that they do contain central BHs (e.g., \citealt{1989ApJ...342L..11F,Reines_2013,Woo_2019,Mezcua_2020}), suggesting that ultra-dwarf galaxies with masses 10$^{5-6}$ \solm, which dominate the number density of galaxies in the Universe, could also harbor such central black holes in the mass range of the GW190521 components.  Attractive evidence for this idea is the fact that these ultra-dwarf galaxies are extremely common, and are even more so in the early universe (e.g., \citealt{Conselice_16}).

Once two ultra-dwarf galaxies merge, it is possible that also the respective central black holes will merge after some time. In fact, this mechanism could be the way SMBHs grow early on in the Universe through hierarchical assembly (e.g. \citealt{Volonteri_assembly}). The question is whether there are enough of these galaxies close to the redshift of the events under consideration, and whether they merge frequently enough, to recover the inferred LIGO/Virgo merger rate for systems such as GW1905121, which is estimated to be  ${0.13}_{-0.11}^{+0.30}\,{{\rm{Gpc}}}^{-3}\,{{\rm{yr}}}^{-1}$ \citep{190521_properties}.

We explore this question in this paper.  In \S 2 we calculate the merger rate for galaxies that could produce an event like GW190521, \S 3 is a discussion of the implications for the proposed formation channel, and in \S 4 we provide a summary and conclusions.

\section{Method}

\begin{figure}
\centering
\includegraphics[width=1\linewidth]{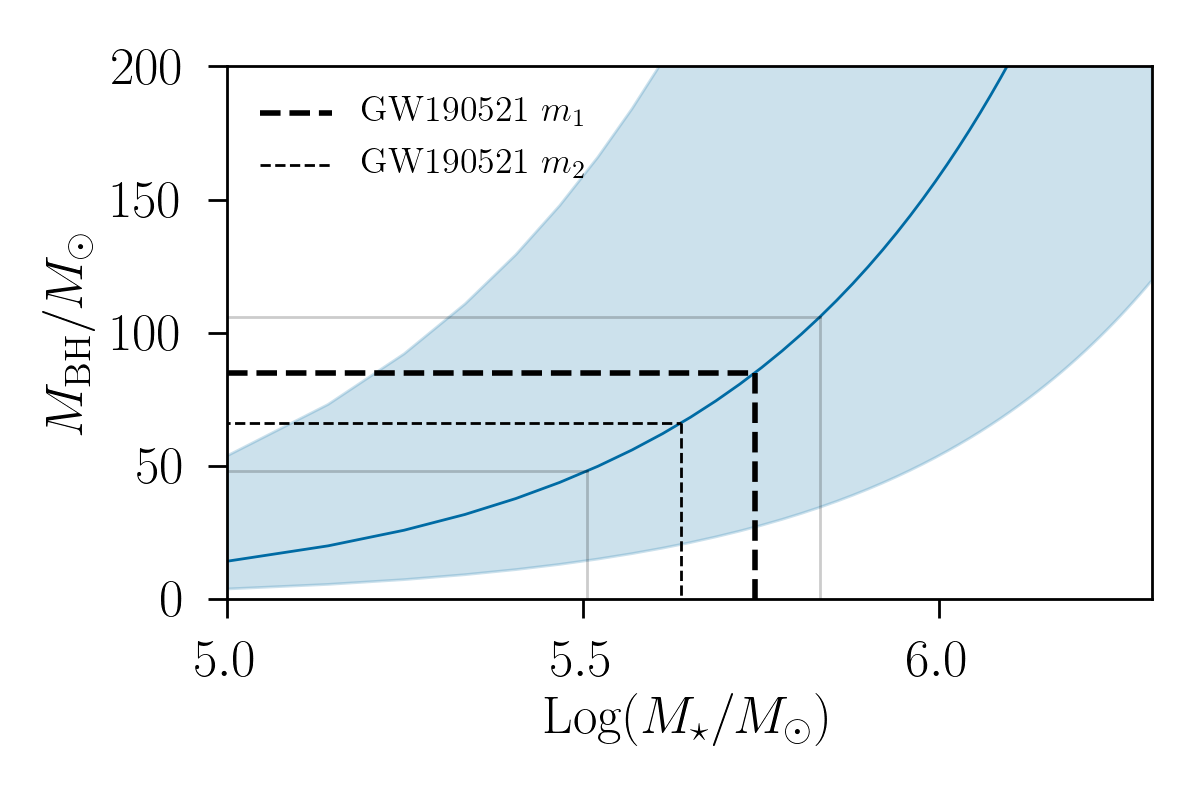}
\caption{Extrapolation to low-masses of the relation between the central black hole mass and total galaxy stellar mass from \cite{Reines_2015}, shown in blue (the shaded region represents the $1\sigma$ uncertainty on the relation's parameters). The masses of the GW190521 components from LIGO/Virgo are shown by the dashed lines. The grey solid lines represent the lowest and highest 90\% CI for both components.}\label{fig:mass}
\end{figure}

To investigate whether GW190521 could be produced through the mergers of central black holes in galaxies we consider the following ingredients: the masses of black holes in extremely low mass galaxies, the merger rate of these galaxies, as well as the time-scales for the black hole mergers to occur after their host galaxies have merged. We update the analysis presented in \citet{Conselice_20}, where we considered all of the LIGO/Virgo mergers from the first two observing runs, to specifically explain GW190521-like events. In order to do this, we first take into account the black hole mass--galaxy mass relation from \citet{Reines_2015}:

\begin{equation}
{\rm log} (M_{\rm BH}) = \alpha + \beta {\rm log} (M_\star/10^{11} {\rm M_{\odot}\,}),\label{eq:mass}
\end{equation}

\noindent where solar mass units are used for the galaxies' stellar mass $M_\star$ and the central BH mass M$_{\rm BH}$. The values of $\alpha$ and $\beta$ are $\alpha = 7.45\pm0.08$ and $\beta = 1.05\pm0.11$. 

Once the mass range of interest is identified based on the masses of the black holes merging, we calculate the merger rate of galaxies in this mass range following \citet{conselice_14}, and describe the volumetric rate $\Gamma_{\rm GM}$ per Mpc$^3$ per Gyr as a function of redshift $z$ as:

\begin{equation}
\Gamma_{\rm GM} (z) = \frac{f(z)}{\tau(z)}{\phi(z)},\label{eq:rate}
\end{equation}

\noindent where $f$ is the fraction of galaxies that merge as a function of redshift, $\phi(z)$ is the number density evolution of the galaxies under consideration, and $\tau (z)$ is the time-scale for galaxy merging, that is how many times do major mergers occur for the population being studied per Gyr.

The best measured major galaxy merger rate as of today is estimated to be close to 0.02 mergers Gyr$^{-1}$, based on merger timescales from \citet{snyder}. However, it is well known that galaxy merging intensifies with lookback time.  The redshift evolution of galaxy mergers is well--described by:

\begin{equation}
f(z) = f_{0} \times (1+z)^{m}
\end{equation}

\noindent where $m$ is the power-law index and $f_{0}$ is the local or $z = 0$ merger fraction for the low-mass galaxies under consideration. Note that this fraction is defined as the number of mergers per galaxy, not the fraction of galaxies merging, which is approximately double the former. Following the findings of \citet{2017MNRAS.470.3507M}, we fix $m = 1.82^{+0.37}_{-0.34}$, and $f_0=0.01^{+0.002}_{-0.002}$ for major mergers of galaxies with similar mass ratios.  

\begin{figure}
\centering
\includegraphics[width=1\linewidth]{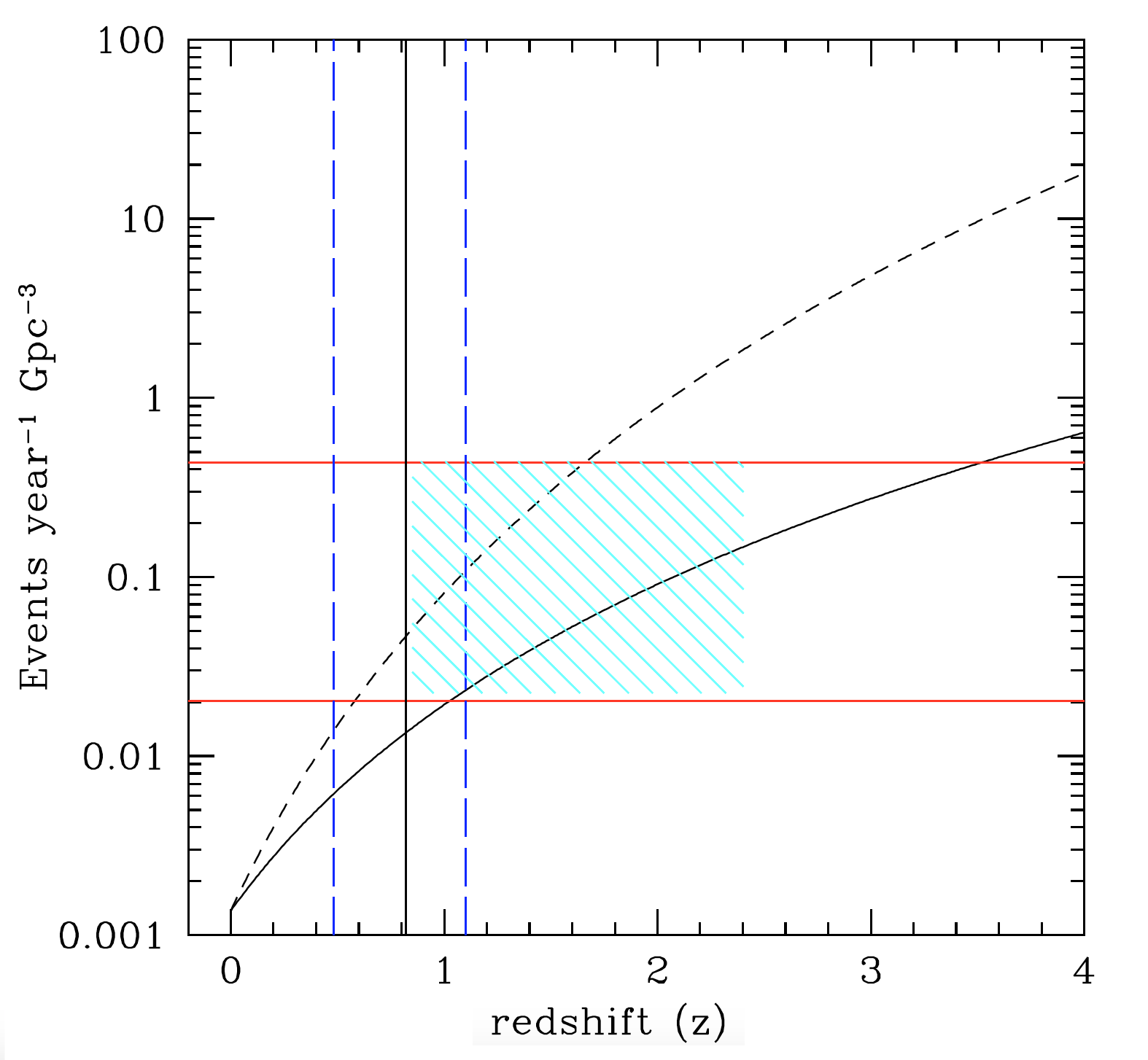}
\caption{Rate of merging ultra--dwarf galaxies in the mass range of interest for GW190521, as a function of redshift. The dashed line is the result using our fit for the number density evolution parameters, while the solid line is the result assuming that the galaxy density is constant with redshift and equal to the one measured at $z=0$. The red lines represent the 90\% CI from the rate estimate of events similar to GW190521 from \citet{GW190521}.  The solid vertical line and blue dashed parallel lines show the range of 90\% CI redshift for GW190521.  The shaded region shows the time-period of 4 Gyr preceding the central redshift.}\label{fig:rate}
\end{figure}

To calculate the number densities of dwarf galaxies we use the results presented in \citet{Conselice_16}, who carried out a compilation of stellar mass functions up to $z \sim 6$ using several observational datasets, and created a model for deriving galaxy stellar mass functions as a function of redshift. 

The number density evolution can be represented by a power-law of the form:

\begin{equation}
\phi(z) = \phi_{0} \times (1+z)^{q}\label{eq:density}
\end{equation}

\noindent where $\phi_0$ is the local or $z = 0$ number density of galaxies in the mass range of interest.   The values we find are: $q = 2.47\pm 0.02$, and $\phi_{0} = 0.086 \pm 0.003$ for low mass galaxies, as explained in Conselice et al. (2020).  However, we renormalize this for the number density of galaxies which map onto the GW190521 system.

At last, the galaxy merging timescale is assumed to follow the relation found by \citet{snyder}: $\tau(z)\propto (1+z)^{-2}$, but with slightly different fits given by a reanalysis of these values presented in Conselice et al. (2020, in prep), such that the time-scale change with redshift given by: $\tau(z) = \tau_{0} \times (1+z)^{u}$.  Combining these, we can calculate the merger rate, in units of per Gyr and per Gpc$^{3}$ given by:

\begin{equation}
\Gamma_{\rm GM} (z)\,  = \frac{f_{0} \phi_{0}}{\tau_{0}} (1+z)^{(m+q+u)}.
\end{equation}

\noindent We therefore use this equation to measure the major merger rate for galaxies at our specific mass range of interest.  

\section{Results and discussion}

The rate of mergers for systems with masses such as GW190521 is inferred to be ${0.13}_{-0.11}^{+0.30}\,{{\rm{Gpc}}}^{-3}\,{{\rm{yr}}}^{-1}$ in \citet{190521_properties}.  In this section, we compare this value to the expected merger rate of galaxies that could host central black holes with masses similar to those measured for the merging components of GW190521.

Based on the central black hole--stellar mass relation in Eq. (\ref{eq:mass}), we extrapolate the range of possible stellar masses of the galaxies hosting central black holes with masses consistent with the components of GW190521. As it can be seen in Fig. \ref{fig:mass}, these correspond to stellar masses in the range $10^5-10^{6.5}~M_\odot$. Galaxies in this mass range have been observed in the local Universe (\citealt{Dekel_2003}; \citealt*{2020ApJ...893...47D}), even down to $\sim 10^3~M_\odot$ \citep{2014ApJ...786...74F}, and have been studied in simulations (e.g. \citealt{tassis}), which show that they are very abundant.

We then use this mass range to estimate the two parameters entering Eq. (\ref{eq:density}), namely 
$\phi_0$ and $q$, by restricting the galaxies from \citet{Conselice_16} to the stellar mass range of interest, as explained in \S 2.  We also examine the number densities of nearby galaxies at the mass range of interest by integrating the stellar mass function between our mass limits using the mass function from \citet{Baldry}.  We use this as a measurement of the $z = 0$ number density for this mass range of objects. 

The final merger rate evolution for the possible host galaxies of GW190521--like BHs is shown in Figure \ref{fig:rate}. The red lines show the upper and lower 90\% CI limits from \citet{190521_properties}. The dashed line is the result using our fit for the number density evolution parameters, while the solid line shows the expected galaxy merger rate assuming that the galaxy density is constant with redshift and equal to the one measured at $z=0$. In both cases, it is clear that the LIGO/Virgo rate for GW190521--like systems can be recovered at around $1\lesssim z\lesssim 2$. For the case of GW190521, which is at $z=0.82^{+0.28}_{-0.34}$, this implies that the likely time delay (i.e. in this case the time between the galaxy merger and the binary merger) is of the order of $\lesssim 4$ Gyr.

For this scenario to be viable we need to understand if time-scales of the order of $\sim 4$ Gyr are reasonable for the BH mergers after the two galaxies have merged. This is a difficult question to answer as these types of galaxies have not been thoroughly studied yet, both observationally and theoretically.  Even within the highest resolution simulations, it is currently not possible to resolve the full dynamics of black holes within merging galaxies \citep{Volonteri_2020}. Analytical arguments are therefore required to estimate the time that needs to elapse between the galaxy merger and the black hole merger. Using simulations of merging galaxies, \citet{Tamfal_2018} find that the central BHs of dwarf galaxies can merge within a Hubble time or stall, depending on the shape of the dark matter profile. BHs in NFW dark matter profiles are however likely to merge. To the best of our knowledge, the simulation in \citet{Tamfal_2018} is the high resolution simulation closest to our case in terms of galaxies' and BHs mass ($\sim10^8~M_\odot $ and $\sim10^5~M_\odot $ respectively, so still larger than the GW190521 case). It is therefore reasonable to consider that the central BHs could also merge in our case if the galaxies have a cuspy dark matter profile. 

Let us assume that the two central BHs sit at the bottom of the host galaxy's gravitational potential well when the two galaxies merge. When the galaxy merger produces a final remnant with a unique core, the central BHs will tend to sink towards the center. If we assume that the BH separation after the remnant is formed is close to $r\sim 80$ pc (note that the typical half right radius of low--mass dwarfs is $\sim 100-400$ pc, e.g. \citealt{Dekel_2003,McConnachie}), then the dynamical friction timescale that will drag the BHs close to the center of the remnant is \citep{binney}:

\begin{equation}
t_{\rm df} = \frac{0.67}{\Lambda} {\rm Gyr} \left(\frac{r}{\rm 4~ kpc} \right)^{2} \left(\frac{\sigma}{\rm 100~ km ~s^{-1}}\right) \left( \frac{\rm 10^{8} M_{\odot}\,}{M} \right)
\end{equation}

\noindent where $\sigma$ is the central velocity dispersion of the galaxy, $\Lambda={\rm ln}(1+M_\star/M)$,  $M$ is the mass of the black hole (or of the star cluster, as we shall assume later), and $M_\star$ is the stellar mass of the remnant galaxy.  Using typical values for dwarf galaxies in the nearby universe, we assume $\sigma \sim 10$ \kms, we find $t_{\rm df}\sim 4$ Gyr for the components of GW190521 at $\sim 80$ pc from the center. 

\begin{figure}
\centering
\includegraphics[width=1\linewidth]{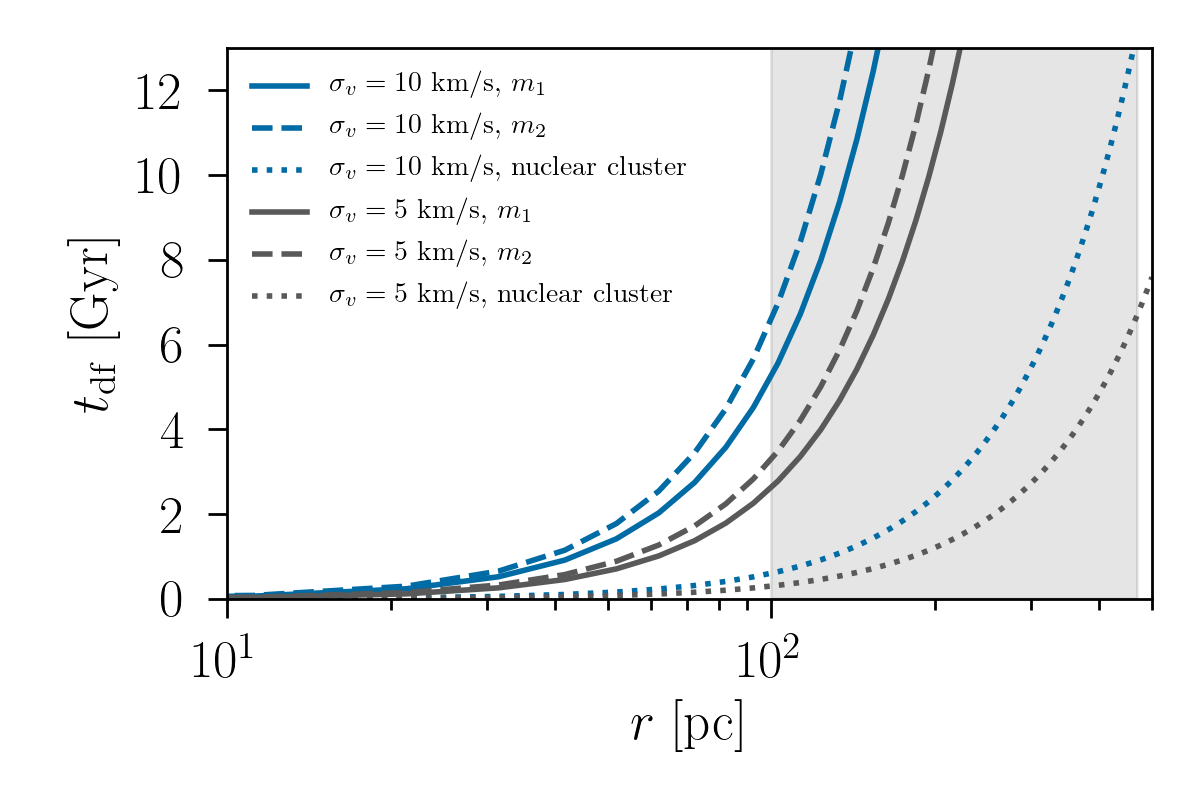}
\caption{Dynamical friction timescale for black holes of mass $m_1=85~M_\odot$ and $m_1=66~M_\odot$ at different separations from the center of the galaxy remnant, having stellar mass $M_\star=10^6~M_\odot$ formed by the merger of two ultra--dwarf galaxies. The shaded region represents the typical half--light radius of nearby $M_\star\sim 10^6~M_\odot$ galaxies, and their velocity dispersion ranges between $5-10$ km~s$^{-1}$. }\label{fig:tdf}
\end{figure}

Figure \ref{fig:tdf} shows the dynamical friction timescales for different values of the central velocity dispersion for the different black hole masses taken into account. If the velocity dispersion of the galaxy is as low as $\sigma \sim 5$ \kms~ (which is a reasonable lower limit for the mass range in consideration; see \citealt{McConnachie,simon}), dynamical friction can be effective in $\sim 4$ Gyr from $\gtrsim 100$ pc (grey lines in Figure \ref{fig:tdf}), thus close to the typical half light radius of 10$^6~M_\odot$ galaxies. Moreover, it is likely that the central BHs are embedded in a nuclear star cluster, for which dynamical friction will be more effective. For a cluster of $1000~M_\odot$ (dotted lines in Figure \ref{fig:tdf}), we find that dynamical friction can be effective within $\sim 4$ Gyr from the edges of the remnant galaxy, at $\sim 200-300$ pc.

At shorter separations, once the binary is formed, hardening by stellar encounters and GW radiation will dominate the binary dynamics. \citet{Biava} find that the duration of these latter phases (referred to as the lifetime of the binary) can take a huge range of values, from fraction of Gyr to more than the Hubble time, depending on the characteristics of the galaxy profile. This scatter in the binary lifetime is even more prominent at the lower masses, and we therefore do not attempt to model these stages.

We conclude that it is reasonable that a BH binary could form and merge within a few Gyr of the merger of the host galaxies, if the BHs reach close enough (of the order of $\sim 80-100$ pc) to the bottom of the potential well of the remnant galaxy, or if they are embedded in star clusters, so that dynamical friction is effective, and if the stellar and dark matter distribution satisfy the criteria that have been explored for higher mass galaxies. In the future, it will be interesting to explore binary formation and lifetime using high resolution simulations for the mass range of interest here.  In particular, it will be interesting to explore the issue of black hole occupation for low-mass dwarfs. It may be possible that not all these systems have central black holes (e.g., \citealt{gallo}), which could result in a better match of merger rates for GW190521--like systems at $z > 3$ in the scenario presented here.

We note that the hardening phase of the binary evolution could increase the binary eccentricity through stellar scattering, and this could be an interesting aspect to explore. In fact, it has been noted for GW190521 that the binary could have had an eccentric orbit \citep{romeroshaw2020gw190521,gayathri2020gw190521}. 

Another binary property of interest for various formation scenarios is the spin. Previous measurements of the effective binary spin $\chi_{\rm eff}$ from population studies hinted to a BBH population with randomly aligned spins \citep{BBH_Properties_O2}, posing a challenge for the isolated binary formation scenario. In the case of merging dwarf galaxies, the binaries do not necessarily need to have aligned spins. Alignment could be facilitated in the case where the binary forms a circumbinary disk in the presence of gas \citep{2020arXiv200606647S}, but not in a generic sample of merging central black holes.

If the formation channel proposed here contributes to a fraction of the observed LIGO/Virgo black holes, this could lead to significant improvements for electromagnetic follow-up campaigns, cosmology, galaxy formation, and fundamental physics. In fact, especially in the case of central black holes as dwarf AGNs, an electromagnetic counterpart could be expected, and a binary AGN could be identified through electromagnetic radiation variability (e.g. \citealt{liao2020discovery}), especially for the most nearby events. Merging activity of dwarf galaxies containing AGNs is likely to affect the majority of dwarfs hosting AGNs, and in fact binary dwarf AGN candidates have already been identified in the nearby Universe \citep{Reines_2020}. If a counterpart is found, binary BHs can also enable standard siren measurements of cosmological parameters \citep{schutz,chen17,palmese2019} that are more precise than the case of BBHs without counterparts (\citealt*{darksiren1}; \citealt{Palmese:2020_190814}), and could even enable measurements of the growth of large scale structure \citep{palmese20_pv}.

An important question is how the black holes in GW190521 could form, even in the case they are the central black holes of galaxies. In this scenario, we believe that the possible mechanisms could be similar to those proposed for the formation of SMBHs. One of the two main channels consists in the formation of very massive, early stars, the Pop III stars \citep{bromm,abel}, which could leave behind black hole seeds from tens to hundreds of solar masses, which would then grow through hierarchical mergers and accretion (e.g. \citealt{Volonteri_assembly}). Another modeling scenario is based on gravitational instabilities in self-gravitating gas clouds, that form an initial black hole of mass $\lesssim 20M_\odot$, which can grow through accretion \citep{Begelman}. In the case of GW190521 the BHs could be very similar to the seeds in the first galaxies.
If the scenario proposed here is confirmed to contribute to the rate of observed BBH mergers, it could open a new observational window into the formation of SMBHs and galaxy formation. On the other hand, if this scenario is ruled out, it would also provide interesting information about the growth of SMBHs through hierarchical assembly.

One way of confirming or ruling out this scenario, would be to search for the electromagnetic counterparts to these events in dwarf galaxies, when they occur in the nearby Universe ($z\lesssim0.1$), where catalogs of faint galaxies (e.g. \citealt{tanoglidis2020shadows}) are already available. Another possibility include a comparison to the expected rate evolution, which is likely to grow with redshift as the galaxy merger rate increases, as shown in \citet{Conselice_20}. At last, we expect that the mass function of detected BHs resembles the galaxy mass function, therefore it would be similar to a modified low-mass end of the Schechter function, although slightly altered as different systems may have different typical time delays.

\section{Summary and conclusions}

In this paper we describe a new, relatively unexplored channel for the production of gravitational wave binary black hole events. We discuss how this channel could be related to the gravitational wave event GW190521, a merger of black holes which has produced the most massive black hole remnant found to date, and also the first intermediate mass black hole known. We argue that the binary components could be the central black holes of merging ultra-dwarf galaxies.

A major astrophysical question is how the black hole progenitors formed and were able to merge with the rate inferred to be ${0.13}_{-0.11}^{+0.30}\,{{\rm{Gpc}}}^{-3}\,{{\rm{yr}}}^{-1}$. Using results and data from different works, we find that the merger rate of ultra-dwarf galaxies at $1<z<2$ is compatible with the inferred rate for GW190521-like events, making our scenario a viable possibility. The required time delay for the black hole merger in the case of GW190521 is likely to be $\lesssim 4$ Gyr.
While we show that typical time delays could be in the order of the Gyrs considering dynamical friction arguments, our findings highlight the necessity of realistic simulations for central black holes in merging ultra-dwarf galaxies to provide more stringent constraints on the expected rate of merging black holes. 
Observations of ultra-dwarf galaxies would also help place limits on the presence of massive central star clusters that can facilitate a faster dynamical friction time-scale \citet{Conselice_20}.

We also note that the proposed scenario could be interesting for the case where only one object is a massive stellar mass black hole (i.e. in our scenario, this would be the central black hole of a dwarf), and the secondary in the binary is a lower mass object, and not a central BH. This possibility could be relevant for e.g. GW190814 \citep{190814_paper} and GW190412 \citep{collaboration2020gw190412}, where the secondary has mass of $\sim 2.6$ and $\sim 8 M_\odot$, respectively. We also do not exclude the possibility that the secondary of GW190521 could be of stellar origin and below the mass gap, implying that the primary would have a mass of 
$\sim 113~ M_\odot$ \citep{fishbach2020dont}. The primary could then be the central black hole of a dwarf with mass $\sim 10^6~M_\odot$.

If confirmed, this scenario would open new avenues in gravitational wave follow--up strategies, cosmology, and in particular galaxy formation and evolution. We expect that as future observations of binary black hole mergers by LIGO/Virgo/KAGRA will build a larger sample of IMBHs and possibly shed light on this formation channel as population analyses provide interesting constraints on the rate evolution and the mass function of these systems. 

\acknowledgments

We thank Alex Drlica-Wagner, Risa Wechsler, James Annis, Daniel Holz for very useful discussion,

Work supported by the Fermi National Accelerator Laboratory, managed and operated by Fermi Research Alliance, LLC under Contract No. DE-AC02-07CH11359 with the U.S. Department of Energy. The U.S. Government retains and the publisher, by accepting the article for publication, acknowledges that the U.S. Government retains a non-exclusive, paid-up, irrevocable, world-wide license to publish or reproduce the published form of this manuscript, or allow others to do so, for U.S. Government purposes.

\bibliographystyle{yahapj_arxiv}
\bibliography{references}



\end{document}